\documentclass[letterpaper]{jpconf}
\usepackage{graphicx}
\begin{document}

\title{Summary: Acoustic Detection of EHE Neutrinos}

\author{J. A. Vandenbroucke}
\address{Dept. of Physics, University of California, Berkeley, CA 94720, USA}
\ead{justinav@berkeley.edu}

\begin{abstract}
Neutrino astronomy was initiated primarily to search for TeV to PeV neutrinos from Active Galactic Nuclei, and the optical Cherenkov technique is well suited for this energy range.  Interest has grown recently in detecting EeV neutrinos, particularly the ``cosmogenic'' neutrinos produced during propagation of ultra-high-energy cosmic rays (UHECR) through the microwave background radiation.  These neutrinos could be a powerful tool both to resolve the mystery of the UHECR sources and to test fundamental physics at the $\sim$100 TeV scale.  The optical technique is not cost effective at these energies and newer techniques such as radio and acoustic detection are necessary.  Accelerator experiments have confirmed the production of both types of signals from high-energy showers in various media, and quantitative measurements have confirmed theoretical descriptions of the signal strength, frequency content and pulse shape.  While radio experiments have set the strongest limits so far, the acoustic method could contribute with an entirely independent signal production and detection mechanism and may be more effective at the highest energies.  Efforts are underway to develop the acoustic method in various media around the world, with arrays operating in ocean water at the Bahamas, the UK, and the Mediterranean Sea; detectors prepared for deployment in the South Pole ice in the next year; and ideas for future acoustic detectors in salt domes and on Antarctica's Ross Ice Shelf.  Regardless of which method is individually most sensitive, the best configuration may be to co-deploy arrays to combine the techniques and seek coincident detection of individual neutrino events.
\end{abstract}

\section{Acoustic Neutrino Detection in Water, Ice, and Salt}
Ocean water is the most abundantly available and best understood acoustic medium for neutrino detection studies.  The acoustic properties of the oceans have been mapped out in detail by military and marine science researchers.  The first detailed search for neutrino-like signals from a large undersea array was completed by the SAUND (Study of Acoustic Ultra-high-energy Neutrino Detection) project \cite{Vandenbroucke05} using a military array in the Bahamas.  Lessons learned in signal simulation, online triggering, and background rejection are now being applied to a larger array at the same location in the SAUND-II project.

Acoustic neutrino arrays in water currently have an energy threshold well above the EeV scale.  Although site selection, signal processing and background rejection advances, and optimization of an array specifically for neutrinos could all lower the threshold, other media likely feature an inherently higher signal to noise ratio resulting in a lower neutrino energy threshold.  Ocean-based arrays may be capable of probing topological defect fluxes and other super-EeV models, but seem not to have sufficient sensitivity at the Greisen-Zatsepin-Kuzmin (GZK) energy range.

Underground salt domes extending several km on each side could have long absorption and scattering lengths for both radio and acoustic waves.  The cost of drilling into these domes, however, is prohibitive.  Drilling into ice is less expensive and may be better suited for radio and acoustic as well as optical neutrino detection.  Greenland ice and ice in temperate glaciers is dirty, ruling out the optical Cherenkov technique, and warm, causing it to absorb radio and acoustic waves well.  South Pole ice, on the other hand, is clean and cold.

To compare media as targets for acoustic neutrino detection, several quantities should be considered: background noise, absorption, scattering, and signal strength (Table~\ref{tab:media}).  In the ocean, background noise is site-dependent but is dominated by weather patterns (wind and rain on the surface).  Transient events that constitute a neutrino background include the sounds of crustacean, cetacean, and human activities.  Noise levels and transient backgrounds in salt and ice are currently unknown in the relevant ($\sim$1-100~kHz) band.  Possible noise sources include seasonal human activity at the station, seasonal wind (note that surface sources are somewhat shielded by the firn, which refracts waves back to the surface), and cracking in the ice bulk or at the bedrock interface due to glacial movement.  Seismically (at $\sim$100 Hz), South Pole is measured to be the quietest place on Earth.

\begin{table}[h]
\caption{\label{tab:media}Acoustic properties of several media.  The neutrino-induced pressure signal strength scales with the Gruneisen parameter.}
\begin{center}
\begin{tabular}{lclclclcl}
\br
					&								& Ocean		& South Pole Ice	& Salt	\\
\mr
Temperature			& $T$ ($^\circ$C)					& 15			& -51			& 30		\\
Sound speed			& $v_L$ (m/s)						& 1530		& 3920			& 4560	\\
Volume expansivity		& $\beta$ (10$^{-5}$ K$^{-1}$)			& 25.5		& 12.5			& 11.6	\\
Heat capacity			& $C_P$ (J/kg/K)					& 3900		&1720			& 839	\\
Peak frequency		& $f_p$ (kHz)						& 7.7			& 20				& 42		\\
Gruneisen parameter	& $\gamma \equiv v_L^2 \beta / C_P$	& 0.153		& 1.12			& 2.87	\\
\br
\end{tabular}
\end{center}
\end{table}

In the ocean, absorption is dominated by dissolved salts.  The absorption length depends on the concentration of these salts locally but is typically several km.  In ice, absorption occurs by reorienting the hydrogen atoms in the crystal lattice \cite{PriceTeV}.  This is a temperature-dependent relaxation process, and thus is depth-dependent at South Pole.  Scattering is insignificant in both water and ice.  In salt, the situation is reversed: The large grain size causes relatively large Rayleigh scattering (scattering length from one to several km).  Absorption in salt (due to anharmonic interactions with thermal phonons) is insignificant; the absorption length is 10's of thousands of km.  However, this assumes pure NaCl.  Inclusions and sediment layers, likely present in salt domes, would significantly increase both scattering and absorption.

Absorption in ocean water scales with frequency-squared and thus high frequency components are preferentially attenuated.  In ice, absorption increases with frequency-squared up to a temperature-dependent turnover frequency and is then flat.  The absorption is frequency-independent above 1~kHz in ice below -24$^\circ$~C, so pulse shape is preserved during propagation.

\section{The Scottish Isles}
The ACORNE (Acoustic {C}{O}smic Ray Neutrino Experiment) collaboration has been using a military array on the island of Rona, Scotland \cite{ThompsonTeV}.  They read out 8 hydrophones continuously at 140 kHz for two weeks in December 2005.  They are now studying signal processing and event reconstruction techniques with this data.  Two significant challenges to acoustic neutrino arrays are to calibrate the phase response of the detector chain and also to develop an in situ calibration device that can mimic the cascade pulse shape and radiation pattern.  They have developed a design that can meet both with a string of several transmitters pulsed in phase.  A clean bipolar acoustic pulse can be generated by inputting a more complicated electrical pulse shape determined using an equivalent LRC circuit model.  The group has also improved simulations of Extremely High Energy (EHE) hadronic cascades \cite{SloanArena}.

\section{The Mediterranean Sea}
In addition to the optical Cherenkov telescopes currently under construction, activities are underway to determine the feasibility of acoustic detection of EHE neutrinos in the Mediterranean Sea.  In contrast with the SAUND and ACORNE projects, these efforts benefit from synergy with the optical telescopes.  They can share deployment procedures, mechanical support, and electrical readout systems with the optical telescopes.  If successful they may be able to search for coincident detection of the same neutrino events with both optical and acoustic arrays, although the energy scales may be too different for sufficient overlap.

Combined event reconstruction could be particularly powerful because optical arrays in the Mediterranean are expected to have excellent pointing resolution in both the muon and the cascade channel (scattering limits the resolution at South Pole, dramatically so for cascades).  Depending on array geometry, three-dimensional acoustic arrays may have $\sim$1$^\circ$ pointing resolution.  In addition to superior combined optical-acoustic reconstruction, a Mediterranean site would be able to view the opposite hemisphere with respect to the South Pole.  Full sky coverage is as important for EeV neutrino arrays as for TeV-PeV neutrino arrays, because GZK neutrinos point back to UHECR sources on cosmological scales.

\begin{figure}
\begin{center}
\includegraphics[width=0.6\linewidth]{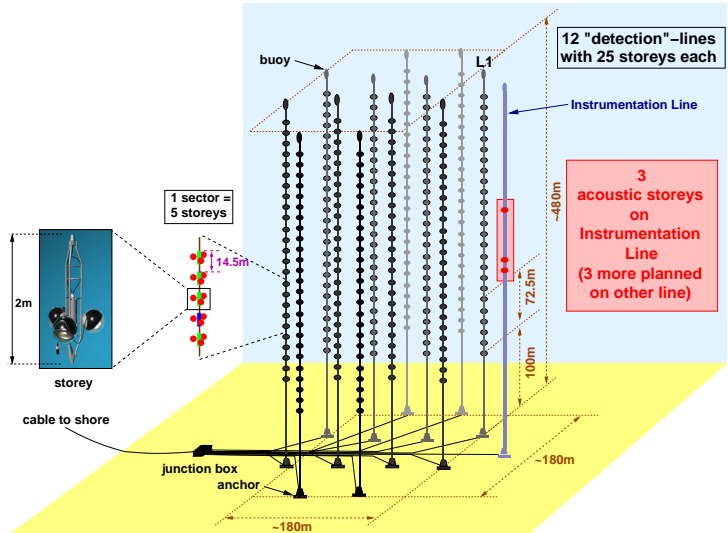}
\end{center}
\caption{\label{fig:antares}The ANTARES detector, showing the instrumentation line that will be deployed in 2007 and includes 3 acoustic storeys comprising AMADEUS.  }
\end{figure}

First, however, sites in the Mediterranean must be evaluated, and signal detection and background reduction techniques developed, for the acoustic technique as well as for the optical.  This is being done as part of both the ANTARES and the NEMO optical neutrino experiments.

In 2005 the ANTARES ``MILOM'' line (mainly a calibration line) was installed with the SPY hydrophone to measure the acoustic background.  In 2007 AMADEUS (ANTARES Modules for Acoustic Detection Under the Sea) \cite{GrafTeV}, a dedicated array of sensors for acoustic studies, will be installed.  AMADEUS (Fig.~\ref{fig:antares}) will comprise three acoustic storeys on the Instrumentation Line, with three more storeys planned for another line.  An acoustic storey is an analog of the optical storey: a $\sim$1-2~m cluster of sensors attached to a line.  On each AMADEUS line, the three acoustic storeys will be spaced vertically 10~m and 100~m apart.  This will allow the study of correlated continuous and transient noise on three distance scales (1, 10, and 100~m).  Two options are being pursued to convert the optical storey design to an acoustic storey design: either six hydrophones will be arranged with three in an upper plane and three in a lower plane within the storey, or the three optical modules of an optical storey will be replaced with three acoustic modules.  Each acoustic module would feature two piezoelectric sensors glued to the inside of a pressure sphere.  The sensors have a self noise level below the sea state zero level in the 1-100~kHz band.

Associated with the NEMO (N{E}utrino Mediterranean Observatory) project off the coast of Sicily, the Ocean Noise Detection Experiment (O$\nu$DE) has been deployed to collect data for 20 months \cite{RiccobeneTeV}.  It consists of a cluster of four hydrophones with a 1-50~kHz band, spaced $\sim$1~m apart, $\sim$2~m off the sea floor at a depth of 2500~m.  Digitized data are routed to shore over optical cable.  It has been running since January 2005 and is characterizing the acoustic environment of the site, which has been shown to have the best optical qualities among investigated sites.

\section{The South Pole}

Given the considerations of background noise, signal strength, absorption, and scattering discussed above, the South Pole is perhaps the most promising location for acoustic neutrino detection.  However, these considerations are based on extrapolations from other measurements.  It remains to determine these quantities in situ.

\begin{figure}
\begin{center}
\includegraphics[width=0.7\linewidth]{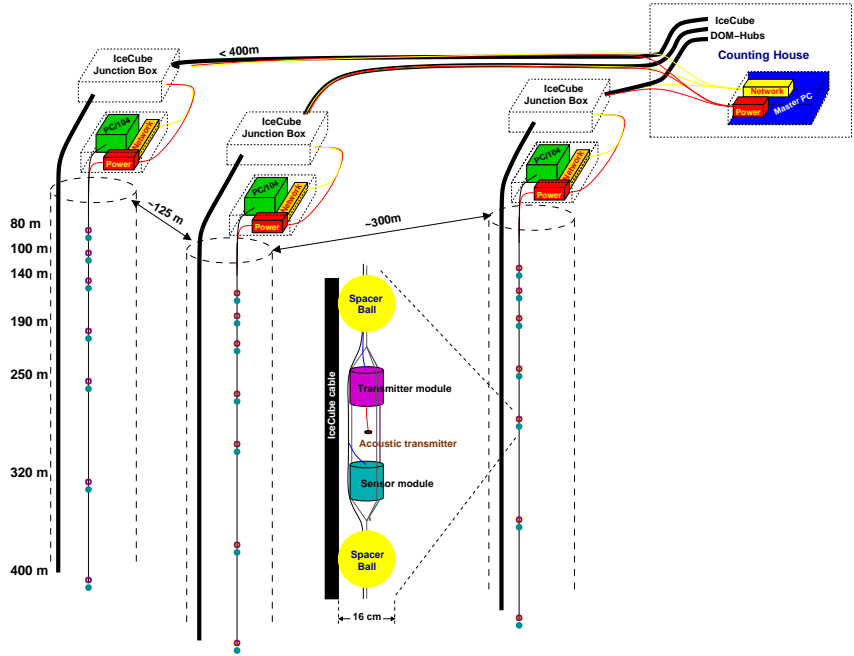}
\end{center}
\caption{\label{fig:spats}The South Pole Acoustic Test Setup.  Three 400~m long strings will be deployed in IceCube holes alongside IceCube strings in January 2007.}
\end{figure}

The South Pole Acoustic Test Setup (SPATS) is designed to do this \cite{HundertmarkTeV}.  It will consist of three 400~m long strings to be deployed in IceCube holes in January 2007 (Fig.~\ref{fig:spats}).  Each SPATS string will be deployed in a hole immediately after the corresponding IceCube string and will be mechanically and electrically independent.   Each string will feature 7 acoustic ``stages,'' with each stage consisting of a transmitter module and a sensor module.  The transmitter module has an epoxy-cast piezoelectric ring that is in direct contact with the ice.  The sensor module has a cylindrical pressure vessel holding three piezoelectric transducers, each pointing in a different azimuthal direction separated by 120$^\circ$.

The absolute uncertainty of both the transmitter amplitude and the sensor sensitivity is $\sim$40\%, and the angular variation in the response of each sensor module is similar.  However, the large channel multiplicity will provide many independent acoustic ray paths.  Combining them will allow a measurement of the absorption length (or a lower limit on it if it is much larger than the SPATS baseline) despite the systematic uncertainty in individual modules \cite{Boeser06}.

\section{Antarctica's Ross Ice Shelf}
Antarctica's Ross Ice Shelf is a body of ice the size of France, several hundred meters thick and floating on sea water.  This unique configuration causes acoustic waves to reflect both off the ice-air interface and (with lower reflectance) off the ice-water interface, effectively trapping some fraction of energy generated by a source in the ice.  Hydrophones have been used to determine that the noise in the water below the ice shelf is much quieter than sea state zero.  Radio waves are reflected entirely off the ice-saltwater interface, making it an interesting site for a radio array as well \cite{BarwickTeV}.

\begin{figure}
\begin{center}
\includegraphics[width=0.6\linewidth]{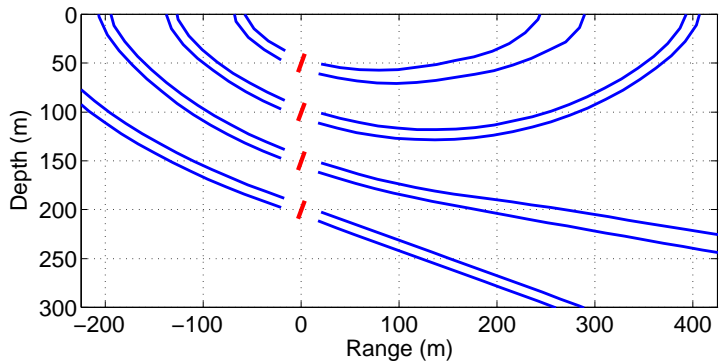}
\end{center}
\caption{\label{fig:rays}Refracted acoustic rays from a neutrino incident from 20$^{\circ}$ zenith and interacting to produce a hadronic cascade at various depths.  Reflections are not shown, but any ray hitting the ice-air interface will reflect with high efficiency.}
\end{figure}

While the acoustic peak frequency emitted by cascades is $\sim$10~kHz, the pulse is broadband.  At frequencies below the peak, the signal spectrum scales with frequency squared.  At these frequencies, the radiation pattern is more isotropic than the flat disk at peak frequency.  For a cascade produced in the upper part of the Ross ice, where the sound speed is increasing with depth as the ice compactifies, some sound rays are refracted up to the surface, reflected, refracted up to the surface again, and so on, effectively propagating as a surface wave (Fig.~\ref{fig:rays}).

There is also the possibility of exciting Rayleigh surface waves.  However, these waves only penetrate a skin depth of order their wavelength, arguing in favor of lower frequency to achieve a greater range of depths in which a cascade could excite such a wave.  Another argument pushing the optimal frequency for a Ross experiment lower is the absorption length:  For shallow South Pole ice (at -51$^\circ$~C), absorptivity scales with frequency-squared up to 30~Hz and is then flat \cite{PriceTeV}.  But for Ross ice (at -28$^\circ$~C), the absorptivity in the flat regime is higher, but the turnover frequency is also higher: 600~Hz.  An experiment optimized below 600~Hz would benefit from the decreased absorption at lower frequencies.

\section{Toward a Hybrid EeV Neutrino Observatory}

There may be a substantial advantage to combining two or three sub-arrays employing different techniques into a single hybrid neutrino detector \cite{VandenbrouckeTeV}.  This is done routinely at particle accelerators, where sub-detectors are optimized for different signals emitted from the same interaction region, and the approach is now the key strategy of the Pierre Auger Observatory in its mission to resolve the UHECR discrepancy.

While the acoustic and radio techniques are promising in the EHE range, they are not well calibrated in comparison with the optical Cherenkov method, which is calibrated using a high rate of atmospheric neutrino events and an even higher rate of atmospheric muons.  The radio technique in particular has been tested with detailed quantitative comparisons between theory and accelerator tests, but except for the small possibility of detecting air shower cores with small optimized arrays, neither will benefit from the in situ calibration sources the optical method enjoys.

However, it may be possible to build a large hybrid array in which both radio and acoustic sub-arrays detect a large GZK rate ($>$10 events per year), with a large fraction ($\sim$half) of events detected in coincidence between the two methods and a small fraction (on the order of 1 event per year) detected in coincidence with an optical array.  Water is good for the optical method, satisfactory for the acoustic method, and impossible for the radio method; salt is good for the radio method, unknown for the acoustic method, and poor for the optical method.  Only ice has the possibility to be well suited for all three techniques.  While the new radio and acoustic ideas for the Ross Ice Shelf are interesting, the prospect of coincident detection with IceCube, the excellent energy and direction reconstruction capabilities of volume arrays compared to surface arrays for all three methods, and the logistical advantages of co-deployment argue strongly in favor of the South Pole.

There are technical questions that must be answered to determine whether this is a viable idea.  Urgent for the acoustic method are the basic material and environmental properties that will be measured by SPATS.  For the radio method, the RICE experiment has already determined that South Pole ice has good radio transparency, but there is still work to be done to optimize the antenna configuration and signal digitization and transmission technology.

Along with recent growth in interest in neutrinos in the EeV range has come a set of new projects aiming to detect these neutrinos acoustically.  Among the three techniques of high energy neutrino detection, the acoustic method involves frequencies five orders of magnitude smaller than the other two, which allows simpler readout technology and makes it the most inexpensive method per channel.  The acoustic method may therefore be the most cost-effective way to study the highest energy range of neutrino astronomy.  Projects now underway are characterizing various media and sites for this purpose as well as developing general techniques that will apply in any medium.  Their results, together with the first astrophysical neutrinos from IceCube and ANITA expected in the near future, will allow us to determine the best way to access the unique physics and astronomy potential of Extremely High Energy neutrinos.\\

\bibliography{references}

\end{document}